\documentclass[11pt]{article}

\usepackage{graphicx}
\usepackage{amsmath}
\usepackage{amsfonts}
\usepackage{amssymb}
\usepackage{verbatim}

\renewcommand{\arraystretch}{1.3}

\catcode`\@=11
\def\marginnote#1{}

\newcount\hour
\newcount\minute
\newtoks\amorpm
\hour=\time\divide\hour by60
\minute=\time{\multiply\hour by60 \global\advance\minute by-\hour}
\edef\standardtime{{\ifnum\hour<12 \global\amorpm={am}%
        \else\global\amorpm={pm}\advance\hour by-12 \fi
        \ifnum\hour=0 \hour=12 \fi
        \number\hour:\ifnum\minute<10 0\fi\number\minute\the\amorpm}}
\edef\militarytime{\number\hour:\ifnum\minute<10 0\fi\number\minute}

\def\draftlabel#1{{\@bsphack\if@filesw {\let\thepage\relax
      \xdef\@gtempa{\write\@auxout{\string
          \newlabel{#1}{{\@currentlabel}{\thepage}}}}}\@gtempa \if@nobreak
    \ifvmode\nobreak\fi\fi\fi\@esphack} \gdef\@eqnlabel{#1}}
    \def\@eqnlabel{}
\def\@vacuum{}
\def\draftmarginnote#1{\marginpar{\raggedright\scriptsize\tt#1}}

\def\draft{
  \oddsidemargin -.5truein
  \def\@oddfoot{\footnotesize \sl preliminary draft \hfil
    \rm\thepage\hfil\sl\today\quad\militarytime}
  \let\@evenfoot\@oddfoot \overfullrule 3pt
    \let\label=\draftlabel
    \let\marginnote=\draftmarginnote
  \def\@eqnnum{(\theequation)\rlap{\kern\marginparsep\tt\@eqnlabel}%
    \global\let\@eqnlabel\@vacuum}

  }

\makeatletter
\newdimen\normalarrayskip              
\newdimen\minarrayskip                 
\normalarrayskip\baselineskip
\minarrayskip\jot
\newif\ifold             \oldtrue            \def\new{\oldfalse}
\def\arraymode{\ifold\relax\else\displaystyle\fi} 
\def\eqnumphantom{\phantom{(\theequation)}}     
\def\@arrayskip{\ifold\baselineskip\z@\lineskip\z@
     \else
     \baselineskip\minarrayskip\lineskip2\minarrayskip\fi}
\def\@arrayclassz{\ifcase \@lastchclass \@acolampacol \or
\@ampacol \or \or \or \@addamp \or
   \@acolampacol \or \@firstampfalse \@acol \fi
\edef\@preamble{\@preamble
  \ifcase \@chnum
     \hfil$\relax\arraymode\@sharp$\hfil
     \or $\relax\arraymode\@sharp$\hfil
     \or \hfil$\relax\arraymode\@sharp$\fi}}
\def\@array[#1]#2{\setbox\@arstrutbox=\hbox{\vrule
     height\arraystretch \ht\strutbox
     depth\arraystretch \dp\strutbox
     width\z@}\@mkpream{#2}\edef\@preamble{\halign
\noexpand\@halignto
\bgroup \tabskip\z@ \@arstrut \@preamble \tabskip\z@ \cr}%
\let\@startpbox\@@startpbox \let\@endpbox\@@endpbox
  \if #1t\vtop \else \if#1b\vbox \else \vcenter \fi\fi
  \bgroup \let\par\relax
  \let\@sharp##\let\protect\relax
  \@arrayskip\@preamble}
\def\eqnarray{\stepcounter{equation}%
              \let\@currentlabel=\theequation
              \global\@eqnswtrue
              \global\@eqcnt\z@
              \tabskip\@centering
              \let\\=\@eqncr

 \halign to \displaywidth\bgroup
    \eqnumphantom\@eqnsel\hskip\@centering
    $\displaystyle \tabskip\z@ {##}$%
    \global\@eqcnt\@ne \hskip 2\arraycolsep
         $\displaystyle\arraymode{##}$\hfil
    \global\@eqcnt\tw@ \hskip 2\arraycolsep
         $\displaystyle\tabskip\z@{##}$\hfil
         \tabskip\@centering
    &{##}\tabskip\z@\cr}
\newfont{\hr}{msbm10}
\newfont{\ams}{msam10}

\hoffset= - 0.9in         

\def\beq{\begin{equation}}
\def\eeq{\end{equation}}
\def\ba{\beq\new\begin{array}{c}}
\def\ea{\end{array}\eeq}
\def\be{\ba}
\def\ee{\ea}
\def\stackreb#1#2{\mathrel{\mathop{#2}\limits_{#1}}}

\def\N2{${\cal N}=2$}

\def\1N{${\cal N}=1$}
\def\4N{${\cal N}=4$}

\newcommand{\rf}[1]{(\ref{#1})}

\newdimen\linethick  \linethick=0.4pt
\newdimen\hboxitspace    \hboxitspace=5pt
\newdimen\vboxitspace    \vboxitspace=5pt

\def\fr#1{%
\beq\new
\vcenter{
\hrule height\linethick
          \hbox{\vrule width\linethick
                \kern\hboxitspace
                \vbox{\kern\vboxitspace
                      \hbox{$\begin{array}{c}\displaystyle#1
         \end{array}$}%
                      \kern\vboxitspace}%
                \kern\hboxitspace
                \vrule width\linethick}%
          \hrule height\linethick}%
\eeq}

\renewcommand{\tt}[1][mer]{\hbox{\tiny{#1}}}

\title{{\bf
On non-conformal limit of the AGT relations} \vspace{.5cm}}
\author{{\bf A. Marshakov}\thanks{E-mail: \ mars@itep.ru; mars@lpi.ru}\ ,\ \
\date{ } 
{\bf A. Mironov}\thanks{E-mail:
\ mironov@itep.ru; mironov@lpi.ru}
\date{ } \\
{\small {\it Theory Department, Lebedev Physics Institute}
and {\it ITEP, Moscow, Russia}}\\ \\
{\bf A.Morozov}\thanks{E-mail: \ morozov@itep.ru}
\date{ } \\ {\small
{\it ITEP, Moscow, Russia}}}

\begin{document}

%

\setcounter{footnote}{3}

\setcounter{tocdepth}{3}

\maketitle

\vspace{-8.cm}

\begin{center}
\hfill FIAN/TD-19/09\\
\hfill ITEP/TH-41/09
\end{center}

\vspace{5.cm}

\begin{abstract}
The Seiberg-Witten prepotentials for ${\cal N}=2$ SUSY gauge theories with
$N_f<2N_c$ fundamental multiplets are obtained from conformal
$N_f=2N_c$ theory by decoupling $2N_c-N_f$ multiplets of heavy matter.
This procedure can be lifted to the level of Nekrasov functions
with arbitrary background parameters $\epsilon_1$ and $\epsilon_2$.
The AGT relations imply that similar limit exists
for conformal blocks (or, for generic $N_c>2$, for the blocks in
conformal theories with $W_{N_c}$
chiral algebra).
We consider the limit of the four-point function explicitly
in the Virasoro case of $N_c=2$, by bringing the dimensions of
external states to infinity.
The calculation is performed entirely in terms of representation theory for
the Virasoro algebra and reproduces the answers
conjectured in arXiv:0908.0307
with the help of the brane-compactification analysis and
computer simulations.
In this limit, the conformal block involving four external primaries, corresponding to the
theory with vanishing beta-function,
turns either into a $2$-point or $3$-point function,
with certain coherent rather than primary external states.
\end{abstract}

\vspace{.5cm}

{\bf 1.} The AGT relations \cite{AGT}-\cite{AGTlast}
express generic $2d$ conformal blocks through the
Nekrasov functions \cite{Nek}-\cite{Neklast} ${\cal Z}(Y)$,
associated with ${\cal N}=2$ SUSY quiver $4d$
gauge theories with extra fundamental multiplets,
generalizing the earlier predictions of \cite{LMN,NO}.
Most commonly these theories
have vanishing beta-functions and
possess conformal invariance in four dimensions.
In the simplest case of the $4$-point Virasoro conformal block, this is the
conformal $SU(2)$ model with $N_f=2N_c = 4$ flavors.
The masses $\mu_1,\ldots,\mu_{N_f}$ of the four fundamentals are
related to the dimensions of four external states operators:
\be
\mu_1 = \alpha_1 - \alpha_2 + \frac{\epsilon}{2}, \ \ \ \
\mu_2=\alpha_1+\alpha_2 - \frac{\epsilon}{2},\ \ \ \
\mu_3 = \alpha_3 - \alpha_4 + \frac{\epsilon}{2}, \ \ \ \
\mu_4=\alpha_3+\alpha_4 - \frac{\epsilon}{2},   \\
\Delta_k = \frac{\alpha_k(\epsilon-\alpha_k)}{\epsilon_1\epsilon_2},
\ \ \ \ \
c=1+\frac{6\epsilon^2}{\epsilon_1\epsilon_2}, \ \ \ \
\epsilon=\epsilon_1+\epsilon_2
\label{muvsdim}
\ee
and the gauge theory condensate (modulus) $a=a_1=-a_2$ is
related to that of the intermediate state:
\be
a = \alpha - {\epsilon\over 2}
\label{avsal}
\ee
For large masses $\mu_k\rightarrow\infty$ the fundamental fields in $4d$ theory decouple,
and one gets an asymptotically free pure gauge ${\cal N}=2$ SUSY theory,
with prepotential expressed through
(the $\epsilon_1=-\epsilon_2\rightarrow 0$ limit of)
the {\it pure gauge} Nekrasov functions $Z(Y)$:
\be
Z(Y) \sim \lim_{\mu_k \rightarrow \infty} {\cal Z}(Y)
\ee
The AGT relation implies that the associated limit of conformal block corresponds to
this $Z(Y)$.
A natural question is how does this limit look like from the point of view
of $2d$ conformal theory itself.

This question was addressed in \cite{Gnc} and an elegant answer has been proposed:
the relevant conformal blocks are matrix elements for certain ``coherent'' states
in Verma module of Virasoro algebra.
However, in \cite{Gnc} the answer was not derived in a {\it direct} way,
by taking a particular limit of the $4$-point conformal block with generic $\mu$'s.
Instead, the conclusion was based
on analysis of the underlying $5$-brane configurations \cite{bran}, which was
also the original source of the AGT relations in \cite{AGT}.
In this letter, we fill the gap and derive the result of \cite{Gnc}
straightforwardly, making use of explicit knowledge of the Virasoro conformal blocks from
\cite{MMMagt}. A similar analysis is possible for conformal blocks with
more external states and for {\it some} $W$-algebra blocks $N_c>2$,
in the last case the results of  \cite{Wyl,MMMM,mmAGT} should be used.
These generalizations are, however, beyond the scope of this paper.

\bigskip

{\bf 2.}
We use notations from \cite{MMMagt} and refer for details and explanations
to that paper.
The $4$-point conformal block is given by the sum over Young diagrams
\be
{\cal B}_{\Delta_1\Delta_2;\Delta_3\Delta_4;\Delta}(x) =
\sum_{|Y|=|Y'|}  x^{|Y|} \gamma_{\Delta\Delta_1\Delta_2}(Y)Q^{-1}_\Delta(Y,Y')
\gamma_{\Delta\Delta_3\Delta_4}(Y')
\label{bqg}
\ee
with the inverse Shapovalov form
$Q_\Delta(Y,Y') = \langle\Delta | L_{Y'}L_{-Y} |\Delta\rangle$,
where
$L_{-Y} = L_{-k_\ell} \ldots L_{-k_2}L_{-k_1}$ for the Young diagram
$Y = \{k_1\geq k_2\geq\ldots\geq k_\ell>0\}$ are made from the Virasoro operators  $L_k$,
$k\in\mathbb{Z}$, satisfying
\be
\label{vialg}
\left[L_m,L_n\right] =
\frac{c}{12}n(n^2-1)\delta_{m+n,0}
+ (m-n)L_{m+n}
\ee
and the three-point functions \cite{Bel,MMMagt} are
\be
\gamma_{\Delta\Delta_1\Delta_2}(Y)
= \prod_{i=1}^{\ell(Y)} \left(\Delta + k_i\Delta_1-\Delta_2+\sum_{j<i}k_j\right)
\label{gammaprod}
\ee
The Shapovalov matrix $Q_\Delta(Y,Y')=Q_\Delta([k_1k_2\ldots],[k'_1k'_2\ldots])$
is infinite-dimensional, but has an obvious block form, since the
matrix elements are non-vanishing only when $|Y|=|Y'|$.
Therefore, for generic $\Delta$ and $c$, it is straightforwardly invertible.
The AGT relations \cite{AGT,MMMagt} state that, under the identification
(\ref{muvsdim}) and (\ref{avsal}),
\be
{\cal B}_{\Delta_1\Delta_2;\Delta_3\Delta_4;\Delta}(x) = {\cal Z}(x)
= \sum_Y x^{|Y|}{\cal Z}(Y)
\ee
and we are going to turn now to the asymptotically free limit of this relation.

\bigskip

{\bf 3.}
We would like to consider first the limit of conformal block (\ref{bqg}), when all
$\mu_1,\ldots,\mu_4 \rightarrow \infty$
independently, and, at the same time, $x\rightarrow 0$ so that
\be
x\prod_{I=1}^4 \mu_I = \Lambda^4
\label{mul}
\ee
which is a scale (like $\Lambda_{QCD}$) parameter,
in the pure \N2 SUSY gauge theory with $N_f=0$.
For this we do not even need an explicit form of the Shapovalov matrix,
since it does not depend on external dimensions $\Delta_{1,2,3,4}$.

However, explicit formula (\ref{gammaprod}) is crucially important.
The number of factors in the r.h.s.
of (\ref{gammaprod}) is equal to the number of rows $\ell(Y)$
(the number of non-vanishing $k$'s) in the Young diagram $Y$,
and it is maximal for fixed $|Y|$ when the diagram consists of a single column, i.e. when
all $k_i=1$, $1\leq i\leq \ell(Y)$ or $\ell(Y)=|Y|$.

Since in our limit $\Delta_i \gg \Delta, 1$, the $\gamma$-factor reduces to
\be
\gamma(Y) \sim \prod_{i=1}^{\ell(Y)} (k_i\Delta_1-\Delta_2)
\label{gpl}
\ee
and of all diagrams of a given size $|Y|$, the sum in (\ref{bqg}) is saturated
by the terms, where $\gamma(Y)$'s (\ref{gpl})
contain maximal possible number of factors, i.e. when $\ell(Y)=|Y|$, or
$Y$ is a single-column diagram
$[1^{|Y|}] = [\underbrace{1,\ldots,1}_{|Y|\ {\rm times}}]$:
\be
x^{|Y|/2}\gamma_{\Delta\Delta_1\Delta_2}(Y) \rightarrow
\Big(\sqrt{x}\,(\Delta_1-\Delta_2)\Big)^{|Y|}\delta(Y,[1^{|Y|}]) =
\\
=
\left(\frac{\sqrt{x}\mu_1\mu_2}{-\epsilon_1\epsilon_2}\right)^{|Y|}\!\!\delta(Y,[1^{|Y|}])
\rightarrow
\left(\frac{\Lambda^2}{-\epsilon_1\epsilon_2}\right)^{|Y|}\!\!\delta(Y,[1^{|Y|}])
\label{gatode} \ee In what follows we often omit the powers of
$-\epsilon_1\epsilon_2$, which can be easily restored from
dimensional consideration. Since the Shapovalov form does not depend
on $\Delta_1,\ldots,\Delta_4$, this means that the limit of
conformal block \be \hspace{-.8cm} \boxed{ B_\Delta(\Lambda) =
\!\!\lim_{\Delta_i\rightarrow\infty} {\cal
B}_{\Delta_1\Delta_2;\Delta_3\Delta_4;\Delta}(x) = \!
\sum_{|Y|=|Y'|} \Lambda^{4|Y|}Q^{-1}_\Delta(Y,Y')
\delta\Big(Y,[1^{|Y|}]\Big)\delta\Big(Y',[1^{|Y|'}]\Big) =\!\sum_n
\Lambda^{4n}Q^{-1}_\Delta([1^n],[1^n]) } \label{Bn} \ee The r.h.s.
of this expression can be treated as a norm (scalar square) of a
peculiar vector in the Virasoro Verma module ${\cal H}_\Delta$ with
the highest weight $\Delta$. Following \cite{Gnc}, we denote it
$|\Lambda^2,\Delta\rangle = \sum_Y C_Y L_{-Y}|\Delta\rangle \in
{\cal H}_\Delta$. Then \be \||\Delta,\Lambda^2\rangle\|^2 = \langle
\Delta,\Lambda^2|\Delta,\Lambda^2\rangle = \sum_{Y,Y'} C_Y
Q(Y,Y')C_{Y'} \ee and, in order to reproduce the r.h.s. of
(\ref{Bn}), one should take
$C_Y=\Lambda^{|Y|}Q^{-1}_\Delta([1^{|Y|}],Y)$, so that \be \boxed{
|\Delta,\Lambda^2\rangle\ = \sum_Y
\Lambda^{2|Y|}Q^{-1}_\Delta\Big([1^{|Y|}],Y\Big)L_{-Y}|\Delta\rangle
} \label{staL} \ee This vector can be characterized as being
orthogonal to all non single-column states $|\Delta,Y\rangle =
L_{-Y}|\Delta\rangle\in {\cal H}_\Delta$ with $Y\neq [1^{|Y|}]$,
since \be \langle\Delta| L_{Y}|\Delta,\Lambda^2\rangle = \sum_{Y'}
\Lambda^{2|Y'|}Q^{-1}_\Delta\Big([1^{|Y'|}],Y'\Big)
\langle\Delta|L_{Y}L_{-Y'}|\Delta\rangle =\\= \sum_{Y'}
\Lambda^{2|Y'|}Q^{-1}_\Delta\Big([1^{|Y'|}],Y'\Big)Q_\Delta(Y',Y) =
\Lambda^{2|Y|}\delta\Big(Y,[1^{|Y|}]\Big) \label{scapro} \ee This
means, in particular, that it is a kind of a ``coherent'' state,
satisfying \be
L_1|\Delta,\Lambda^2\rangle = 
\Lambda^2
|\Delta,\Lambda^2\rangle,   \\
L_k|\Delta,\Lambda^2\rangle = 0, \ \ \ \forall \ \ \ k\geq 2
\label{Nf0}
\ee
The implication (\ref{scapro})$\Rightarrow$(\ref{Nf0}) deserves
more detailed explanation.
Consider the vector $L_k|\Delta,\Lambda^2\rangle\in {\cal H}_\Delta$ for $k>0$.
The coefficients of its expansion over the basis $|\Delta,Y\rangle = L_{-Y}|\Delta\rangle$
in ${\cal H}_\Delta$ are characterized totally by the scalar products
\be
\label{scadu}
\langle \Delta,Y|L_k|\Delta,\Lambda^2\rangle = \langle \Delta|L_YL_k|\Delta,\Lambda^2\rangle =
\sum_{Y'}b^{(k)}_{YY'}\langle \Delta|L_{Y'}|\Delta,\Lambda^2\rangle\
\stackreb{(\ref{scapro})}{=}\
\\
= \sum_{Y'}b^{(k)}_{YY'}\Lambda^{2|Y'|}\delta\Big(Y',[1^{|Y'|}]\Big)
= \sum_{\ell'}b^{(k)}_{Y[1^{\ell'}]}\Lambda^{2\ell'}
\ee
where $\ell'=\ell(Y')=|Y'|$,
i.e. only the Young diagrams $Y' = [1^{|Y'|}]=[1^{\ell(Y')}]$ can contribute.
It is important, however, that due to the Virasoro commutation relations, \rf{vialg}
the sum in \rf{scadu} is restricted by
$|Y'| \leq |Y|+k$ and $\ell(Y')\leq \ell(Y)+1$, meaning that both the
number of boxes in $Y'$ and the number of elementary Virasoro generators in $L_{Y'}$
is less or equal to those in $L_{Y}L_k$; moreover, the structure of Virasoro
algebra \rf{vialg} requires necessarily $k_i(Y') \geq k_j(Y)$, $i,j=1,\ldots,\ell(Y'),\ell(Y)$.
Hence, one gets for \rf{scadu}
\be
\label{scadua}
\langle \Delta,Y|L_k|\Delta,\Lambda^2\rangle
= \sum_{\ell'}b^{(k)}_{Y[1^{\ell'}]}\Lambda^{2\ell'}
= \delta\Big(Y,[1^{\ell}]\Big)\sum_{\ell'\leq \ell+1}b^{(k)}_{[1^{\ell}][1^{\ell'}]}\Lambda^{2\ell'}=
 \delta\Big(Y,[1^{\ell}]\Big)\delta_{k,1}\Lambda^{2\ell+2}
\ee
and this immediately leads to \rf{Nf0}, since for $k>1$ the vector $L_k|\Delta,\Lambda^2\rangle$
is orthogonal to all vectors in ${\cal H}_\Delta$, while for $k=1$ it coincides
with the vector $|\Delta,\Lambda^2\rangle$ up to a numerical factor $\Lambda^2$.

Differently, expanding  $|\Delta,\Lambda^2\rangle =
\sum_{n\geq 0}\Lambda^{2n}|\Delta,n\rangle$, one gets for
\be
\label{deltan}
|\Delta,n\rangle = \sum_{|Y|=n}Q^{-1}_\Delta\Big([1^{n}],Y\Big)L_{-Y}|\Delta\rangle
\ee
that
\be
\label{lkdn}
L_1|\Delta,n\rangle = |\Delta,n-1\rangle,\ \ \ n\geq 0
\\
L_k|\Delta,n\rangle = 0, \ \ \ \forall \ \ \ k\geq 2,\ n\geq 0
\ee
which is exactly the claim of \cite{Gnc}.
Here we have derived and {\it proved} it directly,
taking the limit of the $4$-point conformal block with arbitrary dimensions.
Note that the whole reasoning  is valid for any $\epsilon_1,\epsilon_2$
and $\epsilon$, i.e. for conformal theory with arbitrary central charge $c$ (the
central charge dependence arises in $|\Delta,\Lambda^2\rangle$ through the
inverse matrix of the Shapovalov form).

\bigskip

{\bf 5.} In a similar way, one can consider partial decoupling of the
fundamental matter, corresponding to the models with $N_f=1,2,3$,
when remaining masses (and related combinations of conformal
dimensions) are preserved as free parameters. Let us start with the
case of $N_f=1$. In such a limit, $\mu_{2,3,4}\rightarrow\infty$ with
finite $x\prod_{I=2,3,4}\mu_I=\Lambda_1^3$, but $\mu_1$ remains
finite itself. According to (\ref{muvsdim}), this means that
$\alpha_1$ and $\alpha_2$ go to infinity, but not independently:
their difference remains finite. In terms of conformal dimensions it
means that
\be \Delta_1-\Delta_2 =
\frac{(\alpha_1-\alpha_2)(\alpha_1+\alpha_2-\epsilon)}{-\epsilon_1\epsilon_2}
\sim \frac{(2\mu_1-\epsilon)
\sqrt{\Delta_1}}{\sqrt{-\epsilon_1\epsilon_2}} \ee
i.e. all dimensions are infinite, but
$\frac{\Delta_1-\Delta_2}{\sqrt{\Delta_1}}$ remains finite. Hence,
in this limit the single-column diagrams still dominate to
contribute into $\gamma_{\Delta\Delta_3\Delta_4}(Y)$, like it has
been considered above in the case of flow into pure gauge theory,
but the factor $\gamma_{\Delta\Delta_1\Delta_2}(Y)$ is now dominated
by a different sort of Young diagrams. The reason is that for
$k_i=1$ the factor $k_i\Delta_1-\Delta_2$ turns into
$\Delta_1-\Delta_2$ and grows not as fast as $\Delta_1$ and
$\Delta_2$ themselves. Instead, the dominant contribution comes now
from the Young diagrams of the form $Y=[2^p, 1^q]$ with $|Y|=2p+q$,
$\ell(Y)=p+q$, since for all of them
\be
\gamma_{\Delta\Delta_1\Delta_2}(Y) \sim (2\Delta_1-\Delta_2)^p
(\Delta_1-\Delta_2)^q \sim
\left(\frac{2\mu_1-\epsilon}{\sqrt{-\epsilon_1\epsilon_2}}\right)^q\!
\Delta_1^{p+q/2} \sim \frac{(2\mu_1-\epsilon)^q
(\mu_2/2)^{|Y|}}{(-\epsilon_1\epsilon_2)^{p+q}} \label{gamu}
\ee
Instead of (\ref{Bn}), the limit of conformal block is now given by
(we again omit the powers of $-\epsilon_1\epsilon_2$)
\be
B_\Delta^{N_f=1}(\Lambda_1,m) = \!\!\!\!\!\!\!
\lim_{\stackrel{\Delta_I\rightarrow\infty}{\Delta_1-\Delta_2\sim
2m\sqrt{\Delta_1}}}
{\cal B}_{\Delta_1\Delta_2;\Delta_3\Delta_4;\Delta}(x) =  \\
= \sum_{|Y|=|Y'|}\sum_p (2m)^{|Y|-2p}
\left({x\mu_2\mu_3\mu_4\over 2}\right)^{|Y|} Q^{-1}_\Delta(Y,Y')
\delta\Big(Y,[2^p,1^{|Y|-2p}]\Big)\delta\Big(Y',[1^{|Y|'}]\Big) =
\\
=\sum_{n,p} (2m)^{n-2p}
\left(\frac{\Lambda_1^3}{2}\right)^nQ^{-1}_\Delta
\Big([2^p,1^{n-2p}],[1^n]\Big) =
\langle\Delta,\Lambda_1/2,2m|\Delta,\Lambda_1^2\rangle,
\label{BnNF1}
\ee
where $m=\mu_1-{\epsilon\over 2}$, $\Lambda_1^3=x\mu_2\mu_3\mu_4$ to be fixed when taking
the limit of $x\to 0$ and $\mu_I\to\infty$, $I=2,3,4$, and
\be
\boxed{
|\Delta,\Lambda,m\rangle = \sum_Y \sum_{p}
m^{|Y|-2p}\Lambda^{|Y|}
Q_\Delta^{-1}\Big([2^p,1^{|Y|-2p}],\ Y\Big) L_{-Y}|\Delta\rangle
}
\label{stamL}
\ee
while the vector $|\Delta,\Lambda^2\rangle$ has been already defined in \rf{staL}.
Considering the matrix elements
\be
\label{scamu}
\langle \Delta| L_{Y} |\Delta,\Lambda,m\rangle\ =
\sum_p m^{|Y|-2p}\Lambda^{|Y|}
\delta\Big(Y, [2^p,1^{|Y|-2p}]\Big)
\ee
and
\be
\langle \Delta| L_{Y}L_k |\Delta,\Lambda,m\rangle\ =
\sum_{Y'}b^{(k)}_{YY'}\langle \Delta|L_{Y'}|\Delta,\Lambda,m\rangle\
\stackreb{(\ref{scamu})}{=}\
 \sum_{Y'}b^{(k)}_{YY'}\sum_p m^{|Y'|-2p}\Lambda^{|Y'|}
\delta\Big(Y', [2^p,1^{|Y'|-2p}]\Big)
= \\
=\delta_{k,1}b^{(1)}_{[2^p,1^{|Y|-2p}][2^p,1^{|Y|+1-2p}]}m^{|Y|+1-2p}\Lambda^{|Y|+1}
+\delta_{k,2}b^{(2)}_{[2^p,1^{|Y|-2p}][2^{p+1},1^{|Y|+2-2(p+1)}]}m^{|Y|+2-2(p+1)}
\Lambda^{|Y|+2}
\ee
one proves exactly in the same way as before that
\be
L_1 |\Delta,\Lambda,m\rangle
= m\Lambda|\Delta,\Lambda,m\rangle, \\
L_2|\Delta,\Lambda,m\rangle =
\Lambda^2|\Delta,\Lambda,m\rangle, \\
L_k|\Delta,\Lambda,m\rangle = 0 \ \ \ \ \ {\rm for}\ \ k\geq 3
\label{condNf1}
\ee
again in agreement with the claim of \cite{Gnc}.

Note also that, in the limit when $m\rightarrow \infty$ together
with $\Lambda\rightarrow 0$ so that $m\Lambda = \Lambda_{N_f=0}^2$,
only the term with $p=0$ survives in the sum (\ref{stamL}) and
this state turns into (\ref{staL}):
$|\Delta,m,\Lambda\rangle \rightarrow |\Delta,\Lambda_{N_f=0}^2\rangle$,
while constraints (\ref{condNf1}) turn into (\ref{Nf0}).
It deserves mentioning that, due to separation of powers of $\Lambda_1$ in \rf{BnNF1}
between two vectors in the scalar product (which is, of course, ambiguous),
this limit is a little bit different from
the conventional ``physical'' limit in Seiberg-Witten theory,
$\mu_1\Lambda_1^3\rightarrow \Lambda_{N_f=0}^4$.

The calculation is very similar in the case of $N_f=2$, if keeping finite
the masses $\mu_1$ and $\mu_3$. Then, both the factors
$\gamma_{\Delta\Delta_1\Delta_2}$ and $\gamma_{\Delta\Delta_3\Delta_4}$ behave
according to \rf{gamu}, when taking $\mu_2\to\infty$ and $\mu_4\to\infty$, and
one gets that the conformal block \rf{bqg}
\be
B_\Delta^{N_f=2}(\Lambda_2,m_1,m_3) = \lim_{\stackrel{\Delta_I\rightarrow\infty}
{\stackrel{\Delta_1-\Delta_2\sim
2\mu_1\sqrt{\Delta_1}}{
\Delta_3-\Delta_4\sim
2\mu_3\sqrt{\Delta_3}}}}
{\cal B}_{\Delta_1\Delta_2;\Delta_3\Delta_4;\Delta}(x) =   \\
= \sum_{|Y|=|Y'|}\sum_{p,p'} (2m_1)^{|Y|-2p}(2m_3)^{|Y|-2p'}
\left({\Lambda_2\over 2}\right)^{2|Y|}Q^{-1}_\Delta(Y,Y')
\delta\Big(Y,[2^p,1^{|Y|-2p}]\Big)\delta\Big(Y',[2^{p'},1^{|Y'|-2p'}]\Big) = \\
=\sum_{n,p,p'} (2\mu_1)^{n-2p}(2\mu_3)^{n-2p'}
\left({\Lambda_2\over 2}\right)^{2n}Q^{-1}_\Delta\Big([2^p,1^{n-2p}],[2^{p'},1^{|Y'|-2p'}]\Big) =
\\ =
\langle\Delta,\Lambda_2/2,2m_1|\Delta,\Lambda_2/2,2m_3\rangle
\label{BnNF21}
\ee
in this limit is a scalar product of two states \rf{stamL},
where $m_{1,3}=\mu_{1,3}-{\epsilon\over 2}$, $\Lambda_2^2=x\mu_2\mu_4$, are again to be fixed
finite in the limit of $x\to 0$ and $\mu_{2,4}\to\infty$.

\bigskip

{\bf 6.}
In the case of ``asymmetric limit'', i.e. if
instead of taking $\mu_{2,4}\to\infty$, one decouples, say,
$\mu_{3,4}\to\infty$, no simplification occurs in the factor
$\gamma_{\Delta\Delta_1\Delta_2}(Y)$ in \rf{bqg}, while
the second factor degenerates according to \rf{gatode}, i.e.
$x^{|Y'|}\gamma_{\Delta\Delta_3\Delta_4}(Y') \rightarrow
\Lambda^{2|Y'|}\delta\Big(Y',[1^{|Y'|}]\Big)$.
This means that the conformal block simplifies, though not as drastically
as in the symmetric limit:
\be
{\tilde B}_\Delta^{N_f=2}(\Lambda_2,\mu_1,\mu_2) =
\lim_{\Delta_{3,4}\rightarrow\infty}
{\cal B}_{\Delta_1\Delta_2;\Delta_3\Delta_4;\Delta}(x) =
\sum_Y \Lambda^{2|Y|}
\gamma_{\Delta\Delta_1\Delta_2}(Y)Q^{-1}_\Delta\Big(Y,[1^{|Y|}]\Big)
=  \\
= \langle \Delta,\Lambda^2|V_{\Delta_1}(1)V_{\Delta_2}(0)\rangle
\label{BNf22}
\ee
since \cite{MMMagt}\footnote{See also \cite{MMMM} for detailed discussion of this notion
which becomes quite nontrivial, when going beyond the Virasoro case.}
\be
\gamma_{\Delta\Delta_1\Delta_2}(Y) =
\langle L_{-Y} V_\Delta|V_{\Delta_1}(1)V_{\Delta_2}(0)\rangle
\ee
Thus, the $4$-point conformal block in this limit reduces to a triple
vertex, as was conjectured in \cite{Gnc}. It depends on $\mu_1$
and $\mu_2$ through $\Delta_1$ and $\Delta_2$.

Similarly, if only one mass, say, $\mu_4\rightarrow\infty$, one obtains
\be
B_\Delta^{N_f=3}(\Lambda_3,\mu_1,\mu_2,\mu_3) =
\lim_{\stackrel{\Delta_{3,4}\rightarrow\infty}{
\Delta_3-\Delta_4\sim
2\mu_3\sqrt{\Delta_3}}}
{\cal B}_{\Delta_1\Delta_2;\Delta_3\Delta_4;\Delta}(x) =
\\
=
\sum_Y \sum_p (2\mu_3-\epsilon)^{|Y|-2p}\left({\Lambda\over 2}\right)^{|Y|}
\gamma_{\Delta\Delta_1\Delta_2}(Y)Q^{-1}_\Delta\Big(Y,[2^p,1^{|Y|-2p}]\Big)
=
\\
= \langle \Delta,\Lambda/2,2\mu_3-\epsilon|V_{\Delta_1}(1)V_{\Delta_2}(0)\rangle
\label{BNf3}
\ee
which is again a reduction from the $4$-point function to a $3$-point one.

\bigskip

{\bf 7.}
To conclude, in this paper we have studied the non-conformal limits
(in the sense of $4d$ supersymmetric gauge theory) of
conformal blocks related to Nekrasov partition functions by the AGT
correspondence. We have derived directly from $2d$
CFT analysis the results, conjectured in \cite{Gnc} from brane considerations and
confirmed by computer simulations,
for the asymptotically free limit of conformal blocks.
The proof holds at the level of Nekrasov functions for arbitrary values
of $\epsilon_1,\epsilon_2$ and $\epsilon=\epsilon_1+\epsilon_2$,
the result for the Seiberg-Witten prepotentials \cite{SW} follows \cite{NO,MN}
after taking the limit of $\epsilon_1,\epsilon_2\rightarrow 0$.
The proof is self-consistent within $2d$ CFT, and, in
application to Nekrasov functions, it
assumes that the original AGT relation is correct.
After numerous checks in \cite{AGT}-\cite{AGTlast}
this looks indisputably true,
though so far has been proven exactly \cite{mmNF,mmAGT}
only in the hypergeometric case for the $W$-algebra blocks
with one special, one
fully-degenerate external state and a free field theory like
selection rule imposed on the intermediate state.

There is a number of other interesting limits, which are
natural and well understood from the point of view of $2d$ CFT
(e.g. large intermediate dimension $\Delta$
or the central charge $c$). It can be interesting to find
their interpretation in terms of the Nekrasov functions and/or
instanton expansions in $4d$ SUSY models.

\bigskip

Our work was partly supported by Russian Federal Nuclear Energy
Agency and by the joint grants 09-02-90493-Ukr,
09-02-93105-CNRSL, 09-01-92440-CE. The work of A.Mar. was also supported by
Russian President's Grants of
Support for the Scientific Schools NSh-1615.2008.2, by the RFBR grant 08-01-00667,
and by the Dynasty Foundation. The work of A.Mir. was partly supported by the RFBR grant
07-02-00878, while the work of A.Mor. by the RFBR grant 07-02-00645; the work of A.Mir.
and A.Mor. was also supported by Russian President's Grants of
Support for the Scientific Schools NSh-3035.2008.2 and by the joint program
09-02-91005-ANF.


\begin{thebibliography}{12}

\bibitem{AGT} L.Alday, D.Gaiotto and Y.Tachikawa,
arXiv:0906.3219

\bibitem{Wyl} N.Wyllard, arXiv:0907.2189

\bibitem{MMMagt} A.Marshakov, A.Mironov and A.Morozov, arXiv:0907.3946

\bibitem{Gnc} D.Gaiotto, arXiv:0908.0307

\bibitem{MMMM} Andrey Mironov, Sergey Mironov, Alexei Morozov
and Andrey Morozov,
arXiv:0908.2064

\bibitem{mmNF} A.Mironov and A.Morozov,
arXiv:0908.2190

\bibitem{mmAGT} A.Mironov and A.Morozov, arXiv:0908.2569


\bibitem{IN} S.Iguri and C.Nunez, arXiv:0908.3460

\bibitem{NX} D.Nanopoulos and D.Xie, arXiv:0908.4409

\bibitem{AGTlast1} L.Alday, D.Gaiotto, S.Gukov, Y.Tachikawa and H.Verlinde,
arXiv:0909.0945

\bibitem{AGTlast} N.Drukker, J.Gomis, T.Okuda and J.Teschner,
arXiv:0909.1105

\bibitem{Nek} N.Nekrasov, Adv.Theor.Math.Phys. {\bf 7} (2004) 831-864

\bibitem{FP} R.Flume and R.Pogossian, Int.J.Mod.Phys. {\bf A18} (2003) 2541

\bibitem{LMN}
A.Losev, A.Marshakov and N.Nekrasov, in Ian Kogan memorial volume
{\it From fields to strings: circumnavigating theoretical physics},
581-621; hep-th/0302191

\bibitem{NO}
N.Nekrasov and A.Okounkov, hep-th/0306238

\bibitem{MN} A.Marshakov and N.Nekrasov,
  JHEP {\bf 0701} (2007) 104, hep-th/0612019;\\
  A.Marshakov,
  Theor.Math.Phys.  {\bf 154} (2008) 362
  arXiv:0706.2857;
  arXiv:0810.1536

\bibitem{NY}
H.Nakajima and K.Yoshioka, math/0306198, math/0311058

\bibitem{Shadchin}
S.Shadchin, SIGMA {\bf 2} (2006) 008, hep-th/0601167

\bibitem{Bruzzo}
D.~Bellisai, F.~Fucito, A.~Tanzini and G.~Travaglini,
Phys.\ Lett.\ B {\bf 480} (2000) 365
hep-th/0002110\\
U.Bruzzo, F.Fucito, A.Tanzini, G.Travaglini, Nucl.Phys. {\bf B611}
(2001)
205-226, hep-th/0008225\\
U.Bruzzo, F.Fucito, J.Morales and A.Tanzini, JHEP {\bf 0305} (2003)
054, hep-th/0211108
\\
U.Bruzzo and F.Fucito, Nucl.Phys. {\bf B678} (2004) 638-655,
math-ph/0310036

\bibitem{Neklast}
F.Fucito, J.Morales and R.Pogossian, JHEP, {\bf 10} (2004) 037,
hep-th/040890

\bibitem{bran} E.Witten,
Nucl.Phys., {\bf B500} (1997) 3-42, hep-th/9703166\\
A.Marshakov, M.Martellini and A.Morozov, Phys.Lett.
{\bf B418} (1998) 294-302, hep-th/9706050\\
A.Gorsky, S.Gukov and A.Mironov,
Nucl.Phys., {\bf B517} (1998) 409-461, hep-th/9707120; Nucl.Phys.,
{\bf B518} (1998) 689, hep-th 9710239\\
D.Gaiotto, arXiv:0904.2715\\
Y.Tachikawa, JHEP {\bf 0907} (2009) 067, arXiv:0905.4074\\
D.Nanopoulos and D.Xie, arXiv:0907.1651

\bibitem{Bel} A.Belavin, A.Polyakov, A.Zamolodchikov,
Nucl.Phys., {\bf B241} (1984) 333-380\\
Al.Zamolodchikov and A.Zamolodchikov, {\sl
Conformal field theory and critical phenomena in 2d systems},
2009, 168 p. (in Russian)\\
A.Belavin, private communication

\bibitem{SW}
N.Seiberg and E.Witten,
Nucl.Phys., {\bf B426} (1994) 19-52;
Nucl.Phys., {\bf B431} (1994) 484-550\\
A.Gorsky, I.Krichever, A.Marshakov, A.Mironov, A.Morozov,
Phys.Lett., {\bf B355} (1995) 466-477\\
A.Gorsky, A.Marshakov, A.Mironov, A.Morozov, Phys.Lett., {\bf
B380} (1996) 75-80, hep-th/9603140

\end{thebibliography}
\end{document}